\documentclass[caps,prl,twocolumn,superscriptaddress]{revtex4-2}

\usepackage{array}
\usepackage{graphicx}
\usepackage[utf8]{inputenc}

\usepackage{amsmath}
\usepackage{units}
\usepackage{longtable}

\begin{document}

\title{Bilinear magnetoresistance in HgTe topological insulator: opposite signs at opposite surfaces demonstrated by gate control}


\affiliation{Univ. Grenoble Alpes, CEA, CNRS, SPINTEC, F-38054, Grenoble, France}
\affiliation{Univ. Grenoble Alpes, CNRS, Institut NEEL, F-38042, Grenoble, France}
\affiliation{Univ. Grenoble Alpes, CEA, Leti, F-38000, Grenoble, France}
\affiliation{Univ. Grenoble Alpes, CEA, IRIG-MEM-L\_Sim, F-38000, Grenoble, France}
\affiliation{Unité Mixte de Physique CNRS-Thales, Université Paris-Saclay, 91767 Palaiseau, France}

\author{Yu Fu}
\thanks{These authors contributed equally to this work.}
\affiliation{Univ. Grenoble Alpes, CEA, CNRS, SPINTEC, F-38054, Grenoble, France}

\author{Jing Li}
\thanks{These authors contributed equally to this work.}
\affiliation{Univ. Grenoble Alpes, CEA, Leti, F-38000, Grenoble, France}
\affiliation{Univ. Grenoble Alpes, CEA, IRIG-MEM-L\_Sim, F-38000, Grenoble, France}

\author{Jules Papin}
\thanks{These authors contributed equally to this work.}
\affiliation{Univ. Grenoble Alpes, CNRS, Institut NEEL, F-38042, Grenoble, France}
\affiliation{Univ. Grenoble Alpes, CEA, Leti, F-38000, Grenoble, France}

\author{Paul Noël}
\affiliation{Univ. Grenoble Alpes, CEA, CNRS, SPINTEC, F-38054, Grenoble, France}

\author{Salvatore Teresi}
\affiliation{Univ. Grenoble Alpes, CEA, CNRS, SPINTEC, F-38054, Grenoble, France}

\author{Maxen Cosset-Chéneau}
\affiliation{Univ. Grenoble Alpes, CEA, CNRS, SPINTEC, F-38054, Grenoble, France}

\author{Cécile Grezes}
\affiliation{Univ. Grenoble Alpes, CEA, CNRS, SPINTEC, F-38054, Grenoble, France}

\author{Thomas Guillet}
\affiliation{Univ. Grenoble Alpes, CEA, CNRS, SPINTEC, F-38054, Grenoble, France}

\author{Candice Thomas}
\affiliation{Univ. Grenoble Alpes, CEA, Leti, F-38000, Grenoble, France}

\author{Yann-Michel Niquet}
\affiliation{Univ. Grenoble Alpes, CEA, IRIG-MEM-L\_Sim, F-38000, Grenoble, France}

\author{Philippe Ballet}
\affiliation{Univ. Grenoble Alpes, CEA, Leti, F-38000, Grenoble, France}

\author{Tristan Meunier}
\affiliation{Univ. Grenoble Alpes, CNRS, Institut NEEL, F-38042, Grenoble, France}

\author{Jean-Philippe Attané}
\affiliation{Univ. Grenoble Alpes, CEA, CNRS, SPINTEC, F-38054, Grenoble, France}

\author{Albert Fert}
\affiliation{Unité Mixte de Physique CNRS-Thales, Université Paris-Saclay, 91767 Palaiseau, France}

\author{Laurent Vila}
\affiliation{Univ. Grenoble Alpes, CEA, CNRS, SPINTEC, F-38054, Grenoble, France}

\date{\today}

\begin{abstract}

Spin-orbit effects appearing in topological insulators (TI) and at Rashba interfaces are currently revolutionizing how we can manipulate spins and have led to several newly discovered effects, from spin-charge interconversion and spin-orbit torques to novel magnetoresistance phenomena. In particular, a puzzling magnetoresistance has been evidenced, bilinear in electric and magnetic fields.  Here, we report the observation of bilinear magnetoresistance (BMR) in strained HgTe, a prototypical TI. We show that both the amplitude and sign of this BMR can be tuned by controlling, with an electric gate, the relative proportions of the opposite contributions of opposite surfaces. At magnetic fields of 1 T, the magnetoresistance is of the order of 1 \% and has a larger figure of merit than previously measured TIs. We propose a theoretical model giving a quantitative account of our experimental data. This phenomenon, unique to TI, offers novel opportunities to tune their electrical response for spintronics.\\
\\
\end{abstract}


\maketitle 


Any plane perpendicular to an infinite wire is a symmetry plane. According to Curie's symmetry principle, the electrical resistances of an infinite wire measured for a positive or a negative current are thus identical, \textit{i.e.} $R$ is even with the applied current. Although the application of a magnetic field partially breaks the symmetry of the system,  magnetoresistance (MR) effects, such as the ordinary/Lorentz MR \cite{Gopinadhan2015}, the anisotropic MR \cite{McGuriePotter1975MR3d,BozothPR1946}, or the magnon MR \cite{PhysRevLett.107.136605}, are also even with the magnetic field.  
    
The puzzling discovery of systems in which the measured resistance depends on the sign of the applied current and of the magnetic field has, therefore, attracted lots of attention. The first observation of this unidirectional magnetoresistance was reported in non-magnetic/ferro-magnetic bilayer systems \cite{Avci_2015,Avci2015b} and related to the additional symmetry breaking at the interface. More recently, it has been found that the symmetry breaking by a magnetic field at Rashba interfaces could also result in unidirectional MR, for example in InAs \cite{CHOI2017513}, BiTeBr \cite{Ideue2017}, SrTiO$_{3}$ \cite{He2018STO, Vaz_2020}, and Ge(111) \cite{Guillet2020}. This MR is called Bilinear MR (BMR), as it varies linearly with both the electric current and applied magnetic field. 
In this context, the symmetry breaking at the surface states of three-dimensional (3D) topological insulators (TIs) can lead to BMR as well, as reported for Bi$_{2}$Se$_{3}$ \cite{He2018BiSe}. 

The origin of the BMR in TIs remains debated. Due to substantial spin-orbit coupling, the top and bottom surface states of 3D TIs exhibit spin-momentum locking, \textit{i.e.} the electron spin is locked perpendicularly to the electron momentum. The spin-momentum locking by itself cannot induce a BMR and He \textit{et al.} \ ascribed the BMR of Bi$_{2}$Se$_{3}$ to the hexagonal warping of the band structure \cite{Alpichshev_2010, He2018BiSe}. However, the symmetry of cubic HgTe and (001) surfaces rules out an interpretation of the BMR modeled the hexagonal warping of the band structure. In contrast, Dyrdal \textit{et al.} proposed lately a theoretical model in which the BMR exists with only spin-momentum locking in the absence of hexagonal warping and results very generally from the scattering by inhomogeneities of the spin-momentum locking in the Topological Surface States (TSS) \cite{Dyrdal2020}.

Strained HgTe has been introduced as a topological insulator by Dai et al \cite{Dai_2008}. Here, we report the observation of BMR in strained cubic HgTe 3D topological insulator without hexagonal wrapping. Its figure of merit, $1.2 \times 10^{-3}$ m/(T$\cdot$A), is very large. More importantly, the sign of this BMR can be reversed by applying gate voltages. In our interpretation, we take into account that the opposite surfaces of a TI give rise to opposite BMR, as already predicted by Yasuda et al \cite{Yasuda2016}. We explain the change of sign with gate voltage by the different shifts induced by this voltage in the Fermi levels of the top and bottom surface states. This phenomenon, unique to TI, could foster the introduction of this class of quantum materials in spintronics.

\paragraph{BMR in HgTe}
\begin{figure*}
	\centering
		\includegraphics[width=2.0\columnwidth]{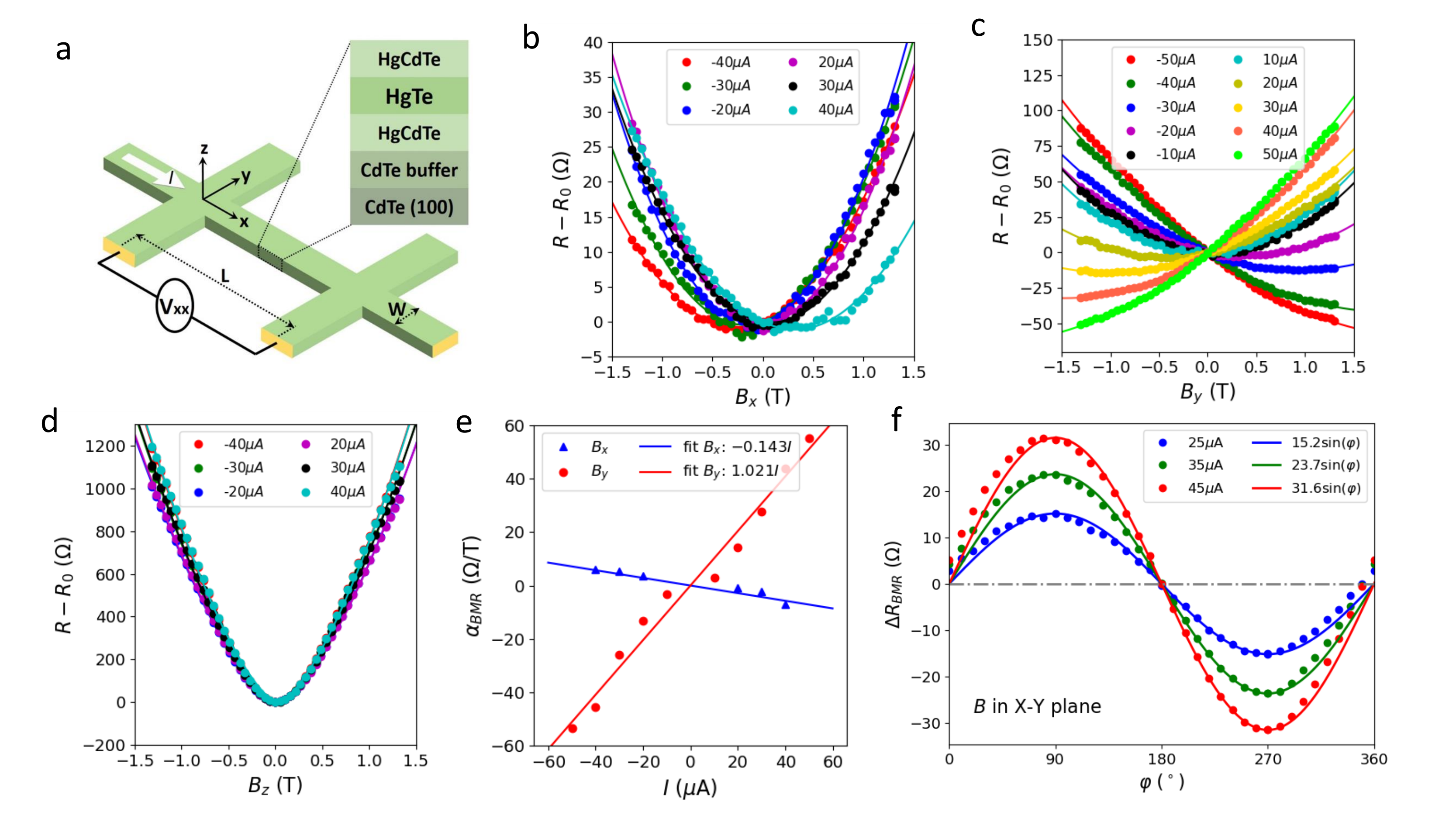}
	\caption{
BMR in HgTe by magnetotransport measurements in Hall bars at $13$ K.    (a): Schematic of Hall bars with width $W=3$ $\mu$m and length $L=15$ $\mu$m. The $30$ nm thick HgTe layer lies between two HgCdTe barriers. A \textit{d.c.} current is injected along the x-axis. 
	(b) to (d): Magnetoresistance versus magnetic field at various \textit{d.c.} currents. Experimental data (dots) follows simple expressions (solid-lines):  $\alpha_{BMR} B_x + \alpha_{Q} B_x^2$ for $B$ along x-axis in (b); $\alpha_{BMR} B_y + \alpha_{Q} B_y^2$ for $B$ along y-axis in (c); and $\alpha_{1.5}|B_z|^{1.5}$ for $B$ along z-axis in (d).
	(e): The coefficient of the linear term to $B$ ($\alpha_{BMR}$) shows a linear dependence on the current $I$, confirming the bi-linearity to $B$ and $I$.
	(f): Azimuthal angular dependence of BMR at $B=0.54$ T at various amplitudes of \textit{d.c.} current. $B$ field lies in X-Y plane, $\varphi =90 ^\circ$ corresponding to $B$ along y-axis. BMR is extracted from the resistances at $\pm I$, \textit{i.e.} $\Delta R_{BMR} = [R(I)-R(-I)]/2$. The experimental data (dots) follow a $\sin(\varphi)$ dependence (solid-lines).
	}
	\label{BMR30nm}
\end{figure*}

We carried out the magneto-transport measurements at low temperature ($13$ K) on a conventional Hall bar device depicted in Fig.~\ref{BMR30nm}a. A \textit{d.c.} current ($I$) is injected along the x axis with the application of an external magnetic field ($B$) ranging from $-1.3$ T to $1.3$ T. The resistance (R) is derived from the measurement of the longitudinal voltage $V_{xx}$. 
Figs.~\ref{BMR30nm}b-\ref{BMR30nm}d report the magnetoresistances ($R-R_0$, where $R_0$ is the resistance at $B=0$ T) for $B$ aligned along x-, y-, and z-axis, respectively. For B along x and z, the magnetoressitance is mainly quadratic with field. For $B$ along $y$ the magnetoresistance follows a simple expression: $\alpha_{BMR} B_y + \alpha_{Q} B_y^2$, which contains two contributions. The first one is odd and linear in field and corresponds to the BMR contribution. The second one is a quadratic contribution, even in field. Furthermore, the coefficient of the linear term ($\alpha_{BMR}$) exhibits a linear dependence on current ($\alpha_{BMR} \propto I$), demonstrated in Fig.~\ref{BMR30nm}e. 

BMR is proportional to the current $I_x$ and magnetic field $B_y$ along the y-axis\cite{He2018BiSe,Dyrdal2020}, \textit{i.e.} $\propto I_x B_y$, and vanishes for $B$ along x- and z-axis. Indeed, we didn't notice any BMR for $B$ along z-axis (Fig.~\ref{BMR30nm}d). 
However, we noticed a weak BMR for $B$ along the x-axis, one order of magnitude smaller than for $B$ along y, as shown in Figs.~\ref{BMR30nm}b and \ref{BMR30nm}e. A misalignment of the sample along x or a current not exactly longitudinal in the sample probably makes that a field along x has a small component perpendicular to the current and induces a small BMR (10 times smaller than with the field along y).
For rotations of the field in the x-y plane (Figs.~\ref{BMR30nm}f), the variation of the BMR as sin($\phi$) agrees with the expected angular dependence \cite{Dyrdal2020} and does not show any detectable deviation which could be related to the strain of the cubic structure. Finally, it can be seen that the magnetoresistance is quasi-isotropic and considerably larger for B along z, see Fig.\ref{BMR30nm}.d which probably involves an additional mechanism of gap opening by an out of plane field. 
We report in SI the same measurements in the Hall configuration to exclude the major role of the thermal effects\cite{Roschewsky_2019,Avci_2014}.

\paragraph{Gate dependence}
\begin{figure*}
	\centering	\includegraphics[width=2.0\columnwidth]{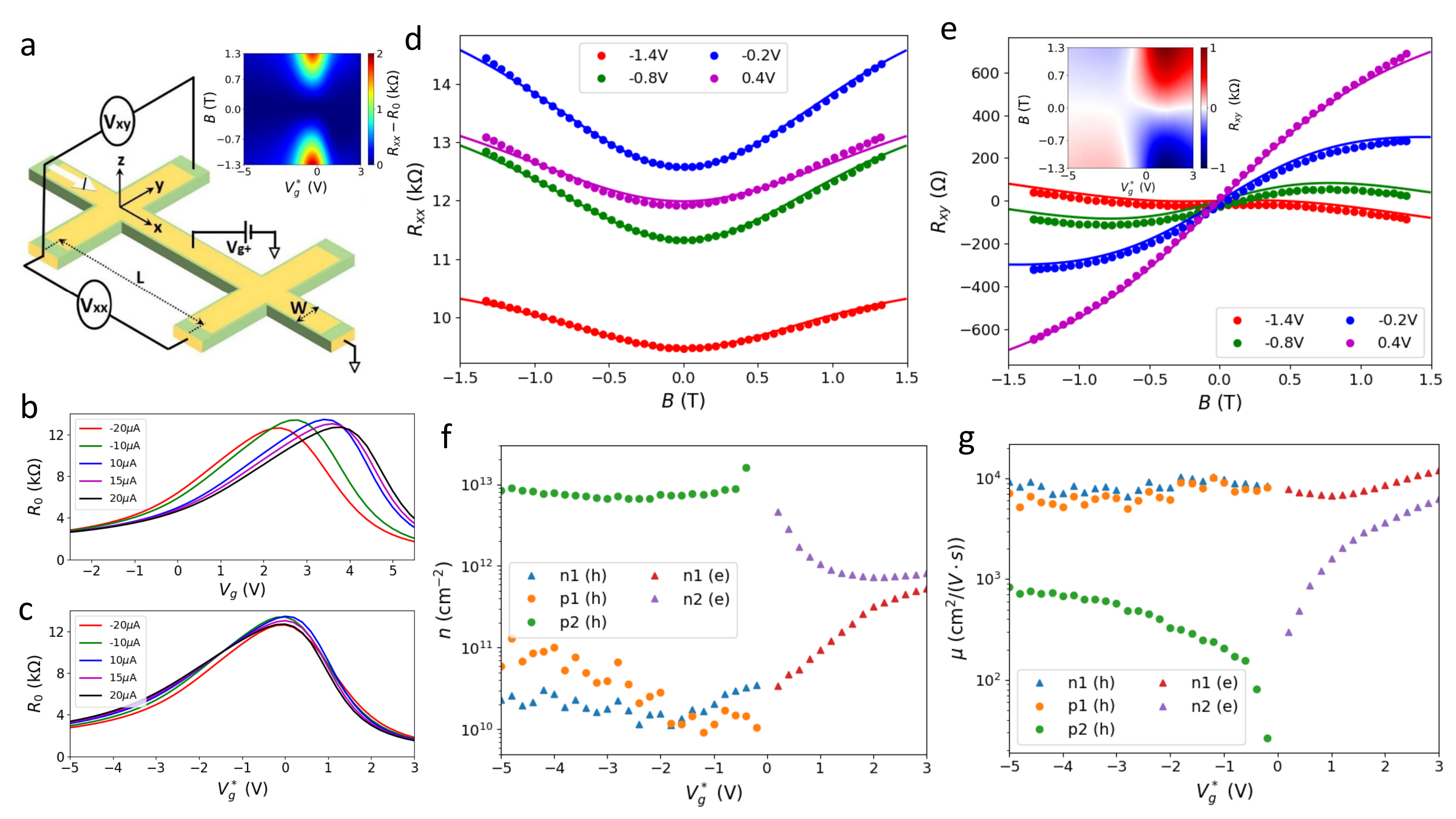}
	\caption{Hall measurement ($B$ along the z-axis) of HgTe Hall bars with an electrical top gate at $13$ K. 
	(a): Schematic of Hall bars with top gate with $W$=3 $\mu$m, $L$=10 $\mu$m. The stacking is similar to Fig.~\ref{BMR30nm}a, but with $20$ nm thick HgTe. Both longitudinal ($V_{xx}$) and lateral ($V_{xy}$) voltages are measured simultaneously. Insert shows the longitudinal resistance versus $V_g^*$ under a field applied along z.
	(b): Longitudinal resistance at $B=0$ T versus gate voltage $V_g$ for various \textit{d.c.} currents. 
	(c): Similar to (b), but for reference gate voltage $V_g^*$. With a rigid shift, the peak resistance is at $V_g^*=0$ V. 
    (d): Longitudinal resistance as a function of the magnetic field for different $V_g^*$ values. Dots correspond to experimental data while solid lines are fits obtained with multi-band modeling. 
    (e): Similar to (d), but for Hall resistance ($R_{xy}=V_{xy}/I_x$). Insert in (e) is the 2D plot of Hall resistance as a function of $V_g^*$ (x-axis), and $B$ (y-axis).  
    (f,g): Carrier densities $n$ and mobilities $\mu$ extracted from Hall measurement at each gate voltages. A 3-bands model (1 electron band + 2 hole bands) is used for $V_g^*\leq 0$, and a 2-bands model (2 electron band) for $V_g^*>0$. The bulk valence band [p2(h)] and bulk conduction band [n2(e)]  have low mobility. The surface electron bands [n1(h) and n1(e)] and surface hole band [p1(h)] have high mobility.
	}
	\label{Hall20nm}
\end{figure*}

To get an insight into the BMR's microscopic origin, we studied devices with a top gate (Fig.~\ref{Hall20nm}a). 
The top gate shifts the Fermi level and impacts charge transport. Fig.~\ref{Hall20nm}b reports the longitudinal resistance without $B$ field ($R_0$) as a function of the gate voltage ($V_g$). 
We define a reference gate voltage ($V_g^*$) identified with respect to the peak resistance of $R_0(V_g)$ curves. Indeed, as the horizontal bias applied to the sample starts to be non-negligible with respect to the top gate voltage, the different current intensities and polarities induce a shift of the electrostatic gate voltage which must be corrected.
As shown by Fig.~\ref{Hall20nm}c, $R_0$ shows a consistent dependence on $V_g^*$ independently of the current intensity. Therefore, $V_g^*$ will be used in the following discussions.

\begin{figure*}
\includegraphics[width=2.0\columnwidth]{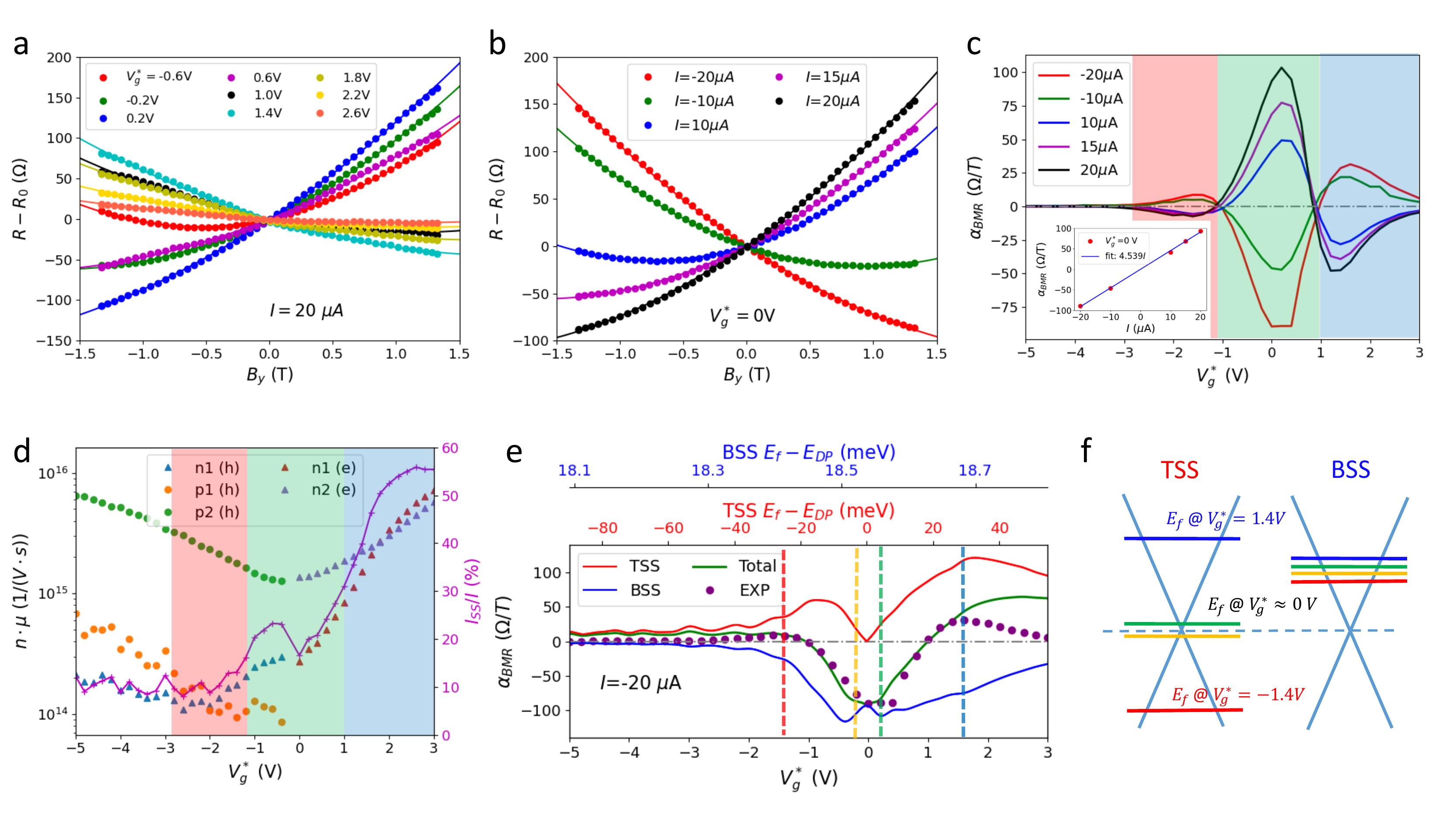}
  \caption{Gate tuned BMR in HgTe at $13$ K. Results are obtained on the same device as in Fig.~\ref{Hall20nm}a.
  (a): Magnetoresistance at various $V_g^*$ for $B$ along y-axis and $I=20$ $\mu$A. 
  (b): Magnetoresistance at various currents for $B$ along y-axis and $V_g^*=0$ V.   
  In both (a) and (b), experimental data (dots) follow $\alpha_{BMR} B_y + \alpha_{Q} B_y^2$ (same as in Fig.~\ref{BMR30nm}c).
  (c): Evolution of the linear term in $B_y$ ($\alpha_{BMR}$) as a function of $V_g^*$ and for different currents. . Insert evidence the linear dependence over the current at $V_g^*=0$ V. The sign of BMR (follows $\alpha_{BMR}$ for positive current) changes twice with gate voltage: positive in green, but negative in red and blue regions.
  (d): Decomposition of conductance into bulk and surface bands according to carrier density and mobility extracted from the Hall measurement (Figs.~\ref{Hall20nm}f and \ref{Hall20nm}g). The ratio of current on surface states ($I_{SS}/I$) is shown by a purple line. The red, green, and blue regions indicate the sign of BMR as in (c).
  (e): The BMR strength for $I = -20\mu$A computed from Eq.~\ref{twosurfaces4}, red (blue) curve for the top (bottom) contributions from the surface states, green curve for the total contribution and dot for the experimental BMR, see text for details. 
  (f): Schematic of the relative positions of the Fermi level with respect to the Dirac point in the calculations from Eq.~\ref{twosurfaces4} leading to Fig.~\ref{BMR20nm}e. In comparison with previous investigation of HgTe topological insulators at much lower temperature in the quantum Hall regime \cite{Ziegler_2020,Brune_2014}, the filling of the bottom surface is here weakly dependent on the gate voltage.
}
\label{BMR20nm}
\end{figure*}

Electric conduction in TIs consists of contributions from bulk and topological surface bands. The peak resistance around $V_g^*=0$ V manifests a minimum contribution from bulk bands, indicating that the Fermi level enters a bulk gap. To separate the bulk and surface transport contributions, we performed Hall measurements at varying gate voltages. The magnetoresistance (Fig.~\ref{Hall20nm}d) and Hall resistance (Fig.~\ref{Hall20nm}e) show that the transition from hole to electron carrier type occurs at $V_g^* \approx 0$ V (the insets in Fig.2a and 2.e show the whole measurements of the magnetoresistance and Hall resistance versus $V_g^*$ respectively).
By using a multi-bands model for Hall measurement (2-bands for $V_g^*>0$ V, and 3-bands model for $V_g^*<0$ V, see methods in detail), we extract the carrier density (Fig.~\ref{Hall20nm}f) and mobility (Fig.~\ref{Hall20nm}g) of the different bands \cite{Kozlov2014}. The two low mobility bands can be identified as the bulk valence band ($V_g^*<0$ V) and the bulk conduction band ($V_g^*>0$ V). The high mobility bands, $\mu\approx10^4$ cm$^2$/(V$\cdot$s), are interpreted as coming from the top and bottom surface bands \cite{Thomas_2017}, with electrons and holes (Fermi energies below and above Dirac Point) for $V_g^*<0$ and only electrons (both top and bottom Fermi energies above Dirac Point) for $V_g^*>0$. At positive $V_g^*$, we cannot distinguish the two high mobility carriers, which are of the same type, so that, only a single high mobility band is identified (sum of the two surface states carrier) in that region.
Therefore, we can decompose the conductance to each band, as shown in Fig.~\ref{BMR20nm}d. As expected, the bulk bands have minimum conductance at $V_g^*=0$ V and the conduction by the surface bands in the middle part of the voltage range is more important.

We performed magneto-transport measurements with $B$ along y-axis at different voltages. As shown in Fig.~\ref{BMR20nm}a, the variation of the resistance as a function of $B_y$ at a given current appears as a combination of linear and quadratic terms and can be fitted as $\Delta R = \alpha_{BMR} B_y + \alpha_{Q} B_y^2$. Fig.~\ref{BMR20nm}b shows that the term $\alpha_{BMR}B_y$ is an odd function of the current and, as already shown in Fig.~\ref{BMR30nm}e, varies linearly with the current as seen in the inset of Fig.~\ref{BMR20nm}.c. The variation of $\alpha_{BMR}$ with the voltage $V_g^*$ is shown in Fig.~\ref{BMR20nm}c, with a positive (negative) maximum for positive (negative) current at around $V_g^*=0$ between two smaller negatives  (positive) maxima at smaller and larger voltages. These successive changes of sign can be explained by the balance between the opposite BMR of the top and bottom interfaces when a gate voltage shifts differently the Fermi levels of the top and bottom surface states with respect to the Dirac Points (DP), as described below.

\paragraph{Interpretation of BMR with opposite contributions from opposite surfaces}
We developed a model on the basis of the theory of the BMR in TIs by Dyrdal et al \cite{Dyrdal2020}. The relative change of conductivity $\sigma$ expected for a single surface state is $\Delta\sigma/\sigma \propto IB_y/|E_f|^3$, where $E_f$ is the Fermi level measured with respect to the DP. This means that a maximum of the BMR is expected at the DP. A 3D TI has opposite surfaces with opposite chiralities and, hence, the BMR amplitudes have opposite signs at the top (t) and bottom (b) surfaces. For current along x-axis, from \cite{Dyrdal2020} we can thus write:
\begin{equation}
\frac{\Delta \sigma_{t(b)}}{\sigma_{t(b)}}\propto +(-)\frac{I_{t(b)}B_y}{|E_{ft(b)}|^3}
\label{twosurfaces1}
\end{equation} 
where $I_{t(b)}$ is the current in the top (bottom) surface state 
\begin{equation}
I_{t(b)}=I_{SS} \frac{\sigma_{t(b)}}{\sigma_t+\sigma_b}
\label{twosurfaces2}
\end{equation}
where $\sigma_t$ and $\sigma_b$ are the conductances of the top and bottom surface states while $I_{SS}$ is the total current in the two surface states. Assuming also that the conductances of the top and bottom surface states are proportional to their respective carrier densities, $\sigma_{t(b)} \propto E_{ft(b)}^2$, the calculation of the BMR-induced magnetoresistance of the two-channel system that we present in the Supplementary Material gives:
\begin{equation}
\alpha_{BMR} \propto \frac{\Delta R}{B_y} \propto I_{SS}\frac{|E_{ft}|}{E_{ft}^2+E_{fb}^2} - I_{SS}\frac{|E_{fb}|}{E_{ft}^2+E_{fb}^2},
\label{twosurfaces3}
\end{equation}
where the first (second) term is the contribution from top (bottom) surface states. Or, it can be expressed as a function of the total (surfaces + bulk) current $I = I_{SS} (\sigma_t + \sigma_b + \sigma_{bulk})/(\sigma_t + \sigma_b)$,
\begin{equation}
\alpha_{BMR} \propto I \frac{\sigma_t + \sigma_b}{\sigma_t + \sigma_b + \sigma_{bulk}} \times \frac{|E_{ft}|-|E_{fb}|}{E_{ft}^2+E_{fb}^2},
\label{twosurfaces4}
\end{equation}
in which the last factor controls the change of sign as a function of the voltage while the first factor accounts for the progressive increase of the proportion of current in the surface states  ($I_{SS}/I$) at large $V_g^*$, as extracted from our measurements (see the purple line in Fig.~\ref{BMR20nm}d).

Eq.~\ref{twosurfaces3} and Eq.~\ref{twosurfaces4} show that, for a given sign of current, the sign of the BMR at a given gate voltage depends on the respective distances, $E_{ft}$ and $E_{fb}$, between the top and bottom Fermi levels and the DP at this voltage. Because the top gate has more impact on the top surface than on the bottom one, we expect $E_{ft}$ to vary faster with $V_g^*$ than $E_{fb}$. 
In Fig.~\ref{BMR20nm}e, we present an example of a good fit we could obtain between the respective voltage dependences of the BMR calculated from Eq.\ref{twosurfaces3} (green line) and the experimental BMR (dots, from measurements at $-20$ $\mu$A). This agreement is obtained by assuming linear variations of $E_{ft}$ and $E_{fb}$ as a function of the voltage, as described in Supplementary Material. 

The main features of the voltage dependence of the BMR in Fig.~\ref{BMR20nm}e can be simply explained by the Fermi level scheme shown in Fig.~\ref{BMR20nm}f in which, as a function of $V_g^*$, the bottom Fermi level $E_{fb}$ remains not too far from the DP while the top Fermi level $E_{ft}$ moves from far below to far above. With $|E_{ft}|\ll |E_{fb}|$ for $V_g^*$ around $0$ V in Fig.~\ref{BMR20nm}f, the predominant contribution to Eq.~\ref{twosurfaces4} comes from the bottom surface, $\alpha_{BMR}\propto -1/|E_{fb}|$, which leads to the negative maximum between $-1$ V and $+1$ V in Fig.~\ref{BMR20nm}e. On both sides of $V_g^* \approx 0$ V, when the top Fermi level is far enough below or above the DP and $|E_{ft}| \gg |E_{fb}|$, Eq.~\ref{twosurfaces4} leads to $\alpha_{BMR}\propto +1/|E_{ft}|$, which explains the positive maxima around $-1.5$ V and $+1.5$ V, larger at $+1.5$ V where the ratio $I_{SS}/I$ is large (see Fig.~\ref{BMR20nm}d). More generally, the progressive increase of $I_{SS}/I$ at large $V_g^*$ (purple line in Fig.~\ref{BMR20nm}d) explains the tendency to higher BMR on the right of Fig.~\ref{BMR20nm}e for both the top and bottom contributions. The temperature dependence (not shown) shows a decrease in the amplitude of the BMR and a less efficient gate effect, the BMR peaks being broader.

\paragraph{Discussion and conclusion}
Our first result is the observation of large BMR effects induced by the surface states of the topological insulator HgTe. The cubic symmetry of HgTe rules out the explanation of BMR by the hexagonal warping put forward in the previous example of BMR in the TI Bi$_2$Se$_3$ \cite{He2018BiSe}. On the contrary, we have shown that our results can be explained by the model of Dyrdal et al \cite{Dyrdal2020}, in which the BMR is induced by the combination of the spin-momentum locking of the surface states of a TI and the scattering by inhomogeneities of this locking. The BMR we observe is particularly large and, in terms of the merit factor $\eta = \Delta R_{BMR}/(R_0BJ)$, is among the largest observed in other experiments with TI or Rashba systems, see Table ~\ref{FOM}.

The second result, unprecedented, is the separation of the opposite BMRs coming from the top and bottom surface of the HgTe layer. It has been obtained by varying a top gate voltage to shift differently the Fermi levels of the top and bottom surface states with respect to their Dirac point. Depending on the gate voltage, the dominant contribution to the parameter $\alpha_{BMR}$ (for $I <0$) comes from the top surface and is positive, or is due to the bottom surface and is negative. To our knowledge, our manuscript presents the first separation of the opposite topological effects on opposite surfaces of a TI and the same type of experiments could be useful to analyze the interplay between opposite interfaces as a function of the thickness.

In addition, a crucial point for the quantitative interpretation of our results is the increase of the BMR of a single surface as $1/|E_f|^3$ at the vicinity of the DP \cite{Dyrdal2020}. This sharp increase overcomes the decrease of the conductance of the surface states as $E_f^2$, which is essential to obtain the behavior described by Eq.~\ref{twosurfaces4} and the fit in Fig.~\ref{BMR20nm}e. This is a good test of the model of Dyrdal et al \cite{Dyrdal2020} since the second existing model \cite{He2018BiSe} predicts the opposite behavior with a BMR tending to zero when the Fermi energy approaches the DP.  

Finally, we believe that our work clears up the problem of the BMR in topological insulators by demonstrating the existence of BMR in symmetry ruling out warping effects, separating the opposite contribution of opposite surfaces to the BMR of a TI, and bringing new data on the variation of the BMR at the vicinity of the Dirac Point.

\begin{table}[t]
  \caption{Figure of merit ($\eta$) for BMR in TI and Rashba systems.}
  \begin{tabular}{lcccc}
  \hline\hline
  Material & System & $\eta$ [$\times 10^{-4}$ m/(T$\cdot$A)] & $T$ (K) & Ref.\\ 
    \hline  
  HgTe & TI    & 12 & 13 & This work \\
  Bi$_{2}$Se$_{3}$ & TI  & 0.001 & 60 & Ref.~\onlinecite{He2018BiSe} \\ 
  BiTeBr & Rashba & 0.15 & 2 & Ref.~\onlinecite{Ideue2017} \\
  SrTiO$_{3}$ & Rashba & 2 & 2 & Ref.~\onlinecite{He2018STO} \\
  Ge(111) & Rashba  & 150 & 15 & Ref.~\onlinecite{Guillet2020}\\
  \hline\hline
  \end{tabular}
  \label{FOM}
\end{table}

\paragraph{Methods}
\label{exp}

Strained HgTe thin layers were grown by molecular beam epitaxy using methods and conditions as described in Refs. \onlinecite{Ballet2014,Thomas2015}. The growth was performed on a (100) CdTe substrate to have HgTe tensile strained, which is essential to turn it into a 3D topological insulator. The growth was initiated with a 200 nm thick CdTe buffer layer to flatten and optimize the substrate surface. The HgTe layer (20 or 30 nm) was deposited in between two Hg$_{0.3}$Cd$_{0.7}$Te(30 nm) barriers to protect against Hg desorption.

The crystal quality was controlled by high-resolution X-ray diffraction \cite{Ballet2014,Thomas2015}. The quality of the interfaces hosting the topological surface states was characterized by scanning transmission electron microscopy \cite{Haas_2017,Noel2018} evidencing an interface width of 1.4 nm.

The HgTe-based heterostructures were then patterned into Hall bars using a low temperature process with standard lithography and evaporation techniques. The fabrication of the top gate comprises the atomic layer deposition of a 25 nm thick Al$_{2}$O$_{3}$ dielectric layer followed by the lithography and evaporation of Ti/Au local gates.  

The devices were then wire-bonded and measured in a temperature variable cryostat equipped with an electromagnet. The current was applied in a continuous mode along the Hall bars, and a constant voltage was applied on the top gate. The longitudinal voltages $V_{xx}$ and $V_{xy}$ were recorded simultaneously using nanovoltmeters. 

Mobility and carrier density are extracted by multi-bands fitting from Hall measurement. Similar to Ref.~\cite{Ali2014}, the conductivity tensor $\sigma$ is expressed in a complex form for a two-bands model (two electron bands):
\begin{equation}
\sigma = e [\frac{n_{e,1} \mu_{e,1}}{1+i\mu_{e,1} B}+\frac{n_{e,2}\mu_{e,2}}{1+i\mu_{e,2} B}],
\end{equation}
and a three-bands model (one electron band and two hole bands):
\begin{equation}
\sigma = e [\frac{n_{e,1} \mu_{e,1}}{1+i\mu_{e,1} B}
+\frac{n_{h,1}\mu_{h,1}}{1-i\mu_{h,1} B}
+\frac{n_{h,2}\mu_{h,2}}{1-i\mu_{h,2} B}].
\end{equation}
The real and imaginary part corresponds to $\sigma_{xx}$ and $\sigma_{xy}$.

\paragraph{Supporting Information}
The Supporting Information is available free of charge at https://pubs.acs.org
The following parts are discussed in the Supporting Information: 1. Development of two surface models for BMR and calculations and 2. Exclusion of Nernst effect.

\paragraph{Author contribution}
 L. V. J.P. A., P. B., T. M. managed the project. J. P., C. T., P. B. prepared the samples. Y. F., J. P., P. N. C. G., M. C-C and T. G. performed the magnetotransport measurements. Y. F., J. P., P. N., S. T. analysed the data. J. L., Y-M N., A. F developed the model. J.P. A, Y. F., J. L., L. V. and A. F. wrote the manuscript. All the authors discussed the data and the manuscript. 

\paragraph{Competing interest}
The authors declare no competing interest.

\paragraph{Data availability}
The data are accessible at the following DOI:https://doi.org/10.57745/ENJADK.
\begin{acknowledgments}
This work was supported by the ANR-Toprise project, the European Union's Horizon 2020 FET-PROACTIVE
project TOCHA under Grant No. 824140 and the Marie Skłodowska-Curie ITN project SPEAR under grant agreement Nº 955671.\\
\\
\end{acknowledgments}


\bibliographystyle{plain}
\bibliography{HgTePaper_v3.4.1}

\end{document}